\documentclass[aps,prc,preprint,amsmath,amssymb,showpacs,preprintnumbers,superscriptaddress]{revtex4}%
\usepackage{multirow}
\usepackage{rotating}
\usepackage{booktabs}
\usepackage{color,graphicx}
\usepackage{amsmath}
\usepackage{dcolumn}
\usepackage{CJK}
\usepackage{bm}
\usepackage{soul}
\usepackage{enumerate}
\usepackage[dvipdfm,bookmarks=true,colorlinks,
citecolor=blue,linkcolor=blue,hypertex,            breaklinks=true]{hyperref}
\usepackage{amsfonts}
\usepackage{amssymb}
\usepackage{graphicx}%
\providecommand{\U}[1]{\protect\rule{.1in}{.1in}}

\begin{document}

\title{Relativistic Brueckner-Hartree-Fock in nuclear matter without the average momentum approximation}
\author{Hui Tong}
\affiliation{State Key Laboratory of Nuclear Physics and Technology, School of Physics, Peking University, Beijing 100871, China}
\author{Xiu-Lei Ren}
\affiliation{State Key Laboratory of Nuclear Physics and Technology, School of Physics, Peking University, Beijing 100871, China}
\affiliation{Institut f\"{u}r Theoretische Physik II, Ruhr-Universit\"{a}t Bochum, D-44780 Bochum, Germany}
\author{Peter Ring}
\affiliation{State Key Laboratory of Nuclear Physics and Technology, School of Physics, Peking University, Beijing 100871, China}
\affiliation{Physik-Department der Technischen Universit\"{a}t M\"{u}nchen, D-85748 Garching, Germany}
\author{Shi-Hang Shen}
\affiliation{State Key Laboratory of Nuclear Physics and Technology, School of Physics, Peking University, Beijing 100871, China}
\author{Si-Bo Wang}
\affiliation{State Key Laboratory of Nuclear Physics and Technology, School of Physics, Peking University, Beijing 100871, China}
\author{Jie Meng\footnote{Email: mengj@pku.edu.cn}}
\affiliation{State Key Laboratory of Nuclear Physics and Technology, School of Physics, Peking University, Beijing 100871, China}
\affiliation{Department of Physics, University of Stellenbosch, Stellenbosch, South Africa}
\affiliation{Yukawa Institute for Theoretical Physics, Kyoto University, Kyoto 606-8502, Japan}
\date{\today}
\begin{abstract}
  Brueckner-Hartree-Fock theory allows to derive the $G$-matrix as an effective interaction between nucleons in the nuclear medium.
  It depends on the center of mass momentum $\bm{P}$ of the two particles and on the two relative momenta $\bm{q}$ and $\bm{q'}$ before and after the scattering process.
  In the evaluation of the total energy per particle in nuclear matter usually the angle averaged center of mass momentum approximation has been used.
  We derive in detail the exact expressions of the angular integrations of the momentum $\bm{P}$ within relativistic Brueckner-Hartree-Fock (RBHF) theory, especially for the case of asymmetric nuclear matter.
  In order to assess the reliability of the conventional average momentum approximation for the binding energy, the saturation properties of symmetric and asymmetric nuclear matter are systematically investigated based on the realistic Bonn nucleon-nucleon potential.
  It is found that the exact treatment of the center of mass momentum leads to non-negligible contributions to the higher order physical quantities.
  The correlation between the symmetry energy $E_{\mathrm{sym}}$, the slope parameter $L$, and the curvature $K_{\mathrm{sym}}$ of the symmetry energy are investigated.
  The results of our RBHF calculations for the bulk parameters characterizing the equation of state are compared with recent constraints extracted from giant monopole resonance and isospin diffusion experiments.
\end{abstract}


\maketitle
\section{Introduction}\label{sec1}

The investigation of the nuclear equation of state (EoS), especially its properties under extreme conditions, is one of the most crucial issues in both nuclear physics and astrophysics.
It is important to understand a variety of interesting phenomena including supernova explosions, the mass-radius correlations of neutron stars, the collective motion of nucleons within the nuclei, the neutron skin thickness of heavy nuclei, as well as some other topics~\cite{LATTIMER2000,Brown2000,Danielewicz2002,Lattimer2004,STEINER2005,BARAN2005,LiChenKo2008,Centelles2009}.
In recent years, with the establishment of many facilities for radioactive ion beams in terrestrial laboratories, such as the Cooling Storage Ring (CSR) Facility in China, the Radioactive Ion Beam (RIB) Factory at RIKEN in Japan, the GSI Facility for Antiproton and Ion Research (FAIR) in Germany, SPIRAL2 at the Grand Accelerateur National d'Ions Lourds GANIL (GANIL) in France, and the Facility for Rare Isotope Beams (FRIB) in the United States, it becomes possible to explore experimentally the EoS of nuclear matter at large isospin asymmetry, in particular, the density dependence of the nuclear symmetry energy.

Theoretical models which are used to investigate the properties of the nuclear EoS can roughly be divided into two methods: Phenomenological and \textit{ab-initio} methods.
Phenomenological methods, either non-relativistic or relativistic, are based on density functionals such as
Skyrme~\cite{Skyrme1956,Vautherin1972_PRC5-626}, Gogny~\cite{Decharge1980_PRC21-1568}, or relativistic mean-field (RMF) models~\cite{Boguta1977_NPA292-413,
Reinhard1989_RPP52-439,RING1996,Meng2006_PPNP57-470,CDFT2016} that are constructed for the purpose to reproduce properties of finite nuclei and nuclear matter. \textit{Ab-initio} methods are based on realistic nucleon-nucleon (NN) interactions with nuclear many-body techniques used for a microscopic treatment of the nuclear system.

There are a variety of formulations of the NN interactions, such as Bonn~\cite{Machleidt1989}, AV18~\cite{Wiringa1995}, CD Bonn~\cite{Machleidt2001}, and chiral potentials~\cite{Kolck2002,Epelbaum2009,MACHLEIDT2011,Xiu-Lei2018}. Recently, more and more \textit{ab-initio} methods have been developed to study the nuclear many-body system, such as the quantum Monte Carlo method~\cite{Carlson2015}, the coupled-cluster method~\cite{Hagen2014}, the no core shell model~\cite{BARRETT2013}, the self-consistent Green's function method~\cite{DICKHOFF2004}, the lattice chiral effective field theory~\cite{LEE2009}, the in-medium similarity renormalization group~\cite{HERGERT2016}, the Monte Carlo shell model~\cite{Otsuka2001_PPNP47-319,Liu2012}, or the Brueckner-Hartree-Fock (BHF) theory~\cite{DAY1967}.
Among these, the relativistic Bonn potential has been successfully applied in relativistic Brueckner-Hartree-Fock (RBHF) theory~\cite{ANASTASIO1983}, to study both nuclear matter~\cite{HOROWITZ1984,Brockmann1990,GQLi1992,Engvik1994,Sehn1997,Jong1998,Alonso2003,Dalen2005,Katayama2013} and, more recently, finite nuclei~\cite{shen2016,Shen2017,SHEN20182,SHEN20181,Shen2018}.

Comparing with non-relativistic BHF, RBHF theory in nuclear matter is relatively complicated and time-consuming.
Therefore, in order to reduce the complexity of this method, in the earlier investigations the so called average center of mass (c.~m.) momentum approximation has been used for the calculation of the binding energy per particle~\cite{Brueckner1968,Alonso2003,Sammarruca2012,Sammarruca2014}.
With the rapid increase of computational power, however, it is now possible to avoid this approximation.
In the present work we derive exact and analytic formulations of the angular integrations for the c.~m. momentum $\bm P$ in the framework of RBHF theory, especially for the asymmetric nuclear matter.
In addition, we systematically study both the density dependence of the energy in symmetric nuclear matter and the symmetry energy at the saturation density $\rho_0$.
For the calculations we  use the Bonn potentials~\cite{Machleidt1989} and compare results with and  without the averaged c.~m. momentum approximation.
In particular we examine the effect of the exact treatment of the c.~m. momentum for the higher order physical quantities in both of the energy in symmetric nuclear matter and the symmetry energy.

In Sec.~\ref{sec2}, we will first describe the general properties of nuclear matter, and then give a brief review of the RBHF framework. Next, we will derive an exact and analytic expression of the angular integrations for the c.~m. momentum $\bm{P}$. Results and discussions are presented in Sec.~\ref{sec3} and a summary is finally given in Sec.~\ref{sec4}.

\section{Theoretical framework}\label{sec2}

\subsection{SATURATION PROPERTIES OF NUCLEAR MATTER}

The binding energy per nucleon of isospin asymmetric nuclear matter can be generally expressed as a power series in the asymmetry parameter $\alpha=(\rho_{n}-\rho_{p})/\rho$, where $\rho=\rho_{n}+\rho_{p}$ is the total density with $\rho_{n}$ and $\rho_{p}$ expressing the neutron and proton densities,
\begin{equation}
  E(\rho,\alpha)=E(\rho,0)+E_{\mathrm{sym}}(\rho)\alpha^{2}+\mathcal{O}(4).
  \label{equ1}
\end{equation}
Here $E(\rho,0)$ is the binding energy per nucleon of symmetric nuclear matter and $E_{\mathrm{sym}}(\rho)$ is the so-called nuclear symmetry energy,
\begin{equation}
  E_{\mathrm{sym}}(\rho)=\frac{1}{2}\frac{\partial^{2}E(\rho,\alpha)}%
  {\partial\alpha^{2}}\bigg|_{\alpha=0}.
  \label{equ2}
\end{equation}

The binding energy per nucleon in symmetric nuclear matter can be expanded around the saturation density $\rho_{0}$,
\begin{equation}
  E(\rho,0)=E(\rho_{0},0)+\frac{K_{\infty}}{2}\left(  \frac{\rho-\rho_{0}}%
  {3\rho_{0}}\right)  ^{2}+\frac{Q_{0}}{6}\left(  \frac{\rho-\rho_{0}}{3\rho
  _{0}}\right)  ^{3}+\mathcal{O}(4),
  \label{equ3}
\end{equation}
where $E(\rho_{0},0)$ denotes the binding energy per nucleon. The second and third derivative of $E(\rho,0)$ with respect to $\rho$ are given by the incompressibility $K(\rho)$ and the skewness parameter $Q(\rho)$,
\begin{align}
  K(\rho)  &  =9\rho_{0}^{2}\frac{\partial^{2}E(\rho,0)}{\partial\rho^{2}},
  \label{equ4}\\
  Q(\rho)  &  =27\rho_{0}^{3}\frac{\partial^{3}E(\rho,0)}{\partial\rho^{3}},
  \label{equ5}
\end{align}
and $K_{\infty}$ and $Q_{0}$ are their values at the saturation density $\rho_{0}$, respectively. The slope of the nuclear matter
incompressibility is given by \cite{Alam2014}%
\begin{equation}
  M(\rho)=3\rho\frac{\partial K(\rho)}{\partial\rho},%
\label{equ6}
\end{equation}
and, at saturation density, we find%
\begin{equation}
  M_{0}=M(\rho_{0})=12K_{\infty}+Q_{0}.
\label{equ7}
\end{equation}
In Ref.~\cite{Alam2016}, the investigation of these quantities shows a strong correlation of the neutron star radii with the slope of the incompressibility.

Similarly, in the vicinity of the saturation density $\rho_{0}$, the symmetry energy can also be characterized in terms of a few bulk parameters:
\begin{equation}
  E_{\mathrm{sym}}(\rho)=E_{\mathrm{sym}}(\rho_{0})+L\left(  \frac{\rho-\rho
  _{0}}{3\rho_{0}}\right)  + \frac{K_{\mathrm{sym}}}{2}\left(  \frac{\rho
  -\rho_{0}}{3\rho_{0}}\right)  ^{2}+\mathcal{O}(3),
  \label{equ8}
\end{equation}
where $E_{\mathrm{sym}}(\rho_{0})$ is the value of the symmetry energy at saturation density, $L$ and $K_{\mathrm{sym}}$ are the slope parameter and curvature parameter of the nuclear symmetry energy at $\rho_{0}$:
\begin{eqnarray}%
  L&=&3\rho_{0}\frac{\partial E_{\mathrm{sym}}(\rho)}{\partial\rho}%
  \bigg|_{\rho=\rho_{0}},
  \label{equ9}\\
  K_{\mathrm{sym}}&=&9\rho_{0}^{2}\frac{\partial^{2} E_{\mathrm{sym}}(\rho
  )}{\partial\rho^{2}}\bigg|_{\rho=\rho_{0}}.
  \label{equ10}
\end{eqnarray}
The nuclear matter incompressibility $K_\infty$ is not a directly measurable quantity.
Instead, one can also define an incompressibility $K_A$ for a finite nucleus with mass number $A$ by measuring the excitation energy of the isoscalar giant monopole resonance (ISGMR)
\cite{Blaizot1980}
\begin{equation}
  E_{\mathrm{ISGMR}}=\sqrt{\frac{\hbar^{2}K_{A}}{M\langle r^{2}\rangle}},%
  \label{equ11}
\end{equation}
where $M$ is the nucleon mass and $\langle r^{2}\rangle$ is the mean square radius of the ground state.
This incompressibility for finite nuclei can be parameterized by means of a similar expansion to the liquid drop mass formula with the volume, surface, symmetry, and Coulomb terms
\cite{Blaizot1980}:
\begin{equation}
  K_{A}\approx K_{\infty}+K_{\mathrm{surf}}A^{-1/3}+K_{\tau}\alpha
  ^{2}+K_{\mathrm{Coul}}\frac{Z^{2}}{A^{4/3}}.
  \label{equ12}
\end{equation}
The symmetry term $K_{\tau}$ and the Coulomb term $K_{\mathrm{Coul}}$ are related to nuclear matter properties as
\cite{Blaizot1980,H.SagawaZhang2007,Colo2014,GargColo2018}:
\begin{eqnarray}%
  K_{\tau}&=&K_{\mathrm{sym}}-6L-\frac{Q_{0}}{K_{\infty}}L,
  \label{equ13}\\
  K_{\mathrm{Coul}}&=&\frac{3}{5}\frac{e^{2}}{r_{0}}\left(  -8-\frac{Q_{0}%
  }{K_{\infty}}\right),
\label{equ14}
\end{eqnarray}
where $r_{0}$ is the radius constant defined by
\begin{equation}
  r_{0}=\left(  \frac{3}{4\pi\rho_{0}}\right)  ^{1/3}.
  \label{equ15}
\end{equation}
If one uses the parabolic approximation in Eq.~(\ref{equ3}) ($Q_{0}=0$), then $K_{\tau}$ can be simplified to
\begin{equation}
  K_{\tau}\approx K_{\mathrm{asy}}=K_{\mathrm{sym}}-6L.
  \label{equ16}
\end{equation}
This equation has been widely used to characterize the isospin dependence of the incompressibility of asymmetric nuclear matter in
Refs.~\cite{Ban2004,BARAN2005,ChenKoLi2005,BAN2006,
LiChenKo2008,Sun2008,Centelles2009}. Obviously, if the skewness parameter $Q_{0}$ is negligible or the magnitude of the slope parameter $L$ is very small, then the coefficient $K_{\mathrm{asy}}$ could be a good approximation to $K_{\tau}$.
Therefore it is important to study in a microscopic approach how the term $Q_{0}$ affects the value of $K_{\tau}$.

As mentioned before, in this investigation we use RBHF theory. In the following, the concepts of this theory in nuclear matter will be briefly reviewed.

\subsection{RELATIVISTIC BRUECKNER-HARTREE-FOCK THEORY}

To evaluate the in-medium nucleon-nucleon potential, one needs a Dirac spinor which is the solution of the Dirac equation for the description of the single-particle motion in the nuclear medium,
\begin{equation}
  u_\tau(\bm{p},s)=\left(
  \frac{E_\tau^{\ast}(p)+M_\tau^{\ast}}{2E_\tau^{\ast}}\right)^{1/2}
  \binom{1}{\frac{\bm{\sigma\cdot p}}{E_\tau^{\ast}(p)+M_\tau^{\ast}}}\chi_{s}.
  \label{equ17}
\end{equation}
Here $M_\tau^{\ast}=M+U_{S,\tau}$ and
${E_\tau^{\ast}}^{2}(p)={M_\tau^{\ast}}^{2}+\bm{p}^{2}$. $U_{S,\tau}$ denotes the scalar potential.
$\tau$ is the isospin quantum number, and $\chi_{s}$ a Pauli spinor.
The normalization is $u^\dagger_\tau(\bm{p},s)u_\tau(\bm{p},s)=1$.

One of the most widely used equations in the RBHF approach is the Thompson equation \cite{Thompson1970}, which is a relativistic three-dimensional reduction of the Bethe-Salpeter equation \cite{Salpeter1951}.
The in-medium Thompson equation describes the scattering of two nucleons in nuclear matter.
It allows to derive the $G$-matrix as an effective interaction in the medium from the solution of the following equation in the momentum space,
\begin{eqnarray}
  G_{\tau_1\tau_2}(\bm{q}^{\prime},\bm{q}|\bm{P},W_{\tau_1\tau_2})=
  V_{\tau_1\tau_2}(\bm{q}^{\prime},\bm{q})
  &+&\int \frac{d^{3}k}{(2\pi)^3}~V_{\tau_1\tau_2}(\bm{q}^{\prime},\bm{k})
  \notag\\
  &&  \times 
  \frac{Q_{\tau_1\tau_2}(\bm{k},\bm{P})}{W_{\tau_1\tau_2}%
  -E_{\tau_1\tau_2}^{\ast}}G_{\tau_1\tau_2}%
  (\bm{k},\bm{q}|\bm{P},W_{\tau_1\tau_2}),
  \label{equ18}
\end{eqnarray}
where $\tau_1\tau_2$ = $nn$,~$pp$,~or~$np$.
$V_{\tau_1\tau_2}$ denotes a realistic bare nucleon-nucleon interaction~\cite{Machleidt1989} and it is constructed in terms of effective Dirac states (in-medium spinors) as explained in Eq.~(\ref{equ17}).
Eq.~(\ref{equ18}) deviates from the Thompson equation (6) in Ref.~\cite{Brockmann1990} by the factor $M^2/E^2$, because we use the Dirac spinors (\ref{equ17}) normalized according to $u^\dag u =1$, as it is usual in many-body physics (see for instance Serot and Walecka in Ref.~\cite{Serot1986}).
$W_{\tau_1\tau_2}$ is the starting energy, and $E_{\tau_1\tau_2}^{\ast}$ is the total energy of intermediate two-nucleon states.
$\bm{P}$ is the c.~m. momentum, $\bm{q}$, $\bm{q'}$, and $\bm{k}$ are the initial, final, and intermediate relative momenta,
\begin{eqnarray}%
  \bm{P}  &=&\frac{\bm{k_1}+\bm{k_2}}{2},
  \label{equ19}\\
  \bm{k}  &=&\frac{\bm{k_1}-\bm{k_2}}{2}.
  \label{equ20}
\end{eqnarray}
the momenta of the two interacting particles $\bm{k_1}$ and $\bm{k_2}$ in nuclear matter can be expressed in terms of the relative momentum $\bm{k}$ and the c.~m. momentum $\bm{P}$.
The Pauli operator $Q_{\tau_1\tau_2}(\bm{k},\bm{P})$ avoids the scattering into occupied states.
It is defined as
\begin{equation}
  Q_{\tau_1\tau_2}(\bm{k},\bm{P})=%
  \begin{cases}
  1,\qquad & |\bm{P}+\bm{k}|>k_{F}^{\tau_1}~\mathrm{or}~|\bm{P}-\bm{k}|>k_{F}^{\tau_2} \\
  0,\qquad & \mbox{otherwise}.
  \end{cases}
  \label{equ21}
\end{equation}
where $Q_{\tau_1\tau_2}(\bm{k},\bm{P})$ depends not only on the magnitude of the c.~m. and relative momentum but also on their relative direction.
To simplify such an angular dependence, one usually replaces the Pauli operator $Q_{\tau_1\tau_2}(\bm{k},\bm{P})$ by an angle-averaged Pauli operator $Q_{\tau_1\tau_2}^{\mbox{av}}(k,P)$ (see Eq.~(\ref{equA1}) in Appendix~\ref{AppendA}).
Several non-relativistic investigations have been carried out to calculate the nuclear matter properties using the exact Pauli operator $Q_{\tau_1\tau_2}(\bm{k},\bm{P})$, and almost all the results have assessed the reliability of this angle-averaged approximation in the non-relativistic framework \cite{Schiller1999,Schiller1999E,SUZUKI2000}.
Therefore we use this approximation also in the relativistic case.
For asymmetric nuclear matter, this value has to be carefully investigated and the details are given in the Appendix~\ref{AppendA}.

After the solution of Eq.~(\ref{equ18}) for the positive energy solutions, the knowledge of the $G$-matrix allows us to calculate the self energy:%
\begin{equation}
  U_{\tau_1}(m)=\sum_{s_{n},\tau_2}
  \int_{0}^{k_{F}^{\tau_2}}d^{3}k_{n}
  \langle mn|G_{\tau_1\tau_2}(W_{\tau_{1}\tau_{2}})|mn-nm\rangle,
  \label{equ22}%
\end{equation}
for the positive energy solutions.
Here $m$ specifies a state below or above the Fermi surface with momentum $\bm{k}_{m}$ and spin $s_m$.
$W_{\tau_{1}\tau_{2}}$ is the starting energy and we use in the following calculations the "continuous choice"~\cite{Jeukenne1976_PR25-83,Baldo1991_PRC43-2605},
\begin{equation}
  W_{\tau_{1}\tau_{2}}
  =E_{\tau_{1}}^{\ast}(p_{m})+E_{\tau_{1}}^{\ast}(p_{n}).
  \label{equ23}
\end{equation}
Before solving the relativistic Hartree-Fock equations in a self-consistent way, one needs the full relativistic
single-particle potential ${U}(p)$, the full self-energy, i.e. matrix elements not only for the positive energy solutions given in Eq.~(\ref{equ22}), but also the elements coupling positive with negative energy solutions and those for the negative with negative energy solutions.
Following the usual prescriptions~\cite{Brockmann1990,Sammarruca2012}, where the Thompson equation is solved only for the positive energy solutions, and neglecting the space-like component of the vector field because of time-reversal invariance, we use the following ansatz for the single-particle potential:
\begin{equation}
  U(p) = U_S + \gamma_0 U_V.
  \label{equ24}
\end{equation}
Furthermore, the momentum dependence of the scalar and vector fields is very weak and neglected.
The two constants $U_S$ and $U_V$ are adjusted to the positive energy solutions in Eq.~(\ref{equ22}) at the Fermi momentum.
This leads to the relativistic Hartree-Fock equation:
\begin{equation}
  \{\bm{\alpha \cdot p}+U_V+\beta M^\ast\}u(\bm{p}) = E(p) u(\bm{p}),
  \label{equ25}
\end{equation}
where $\bm{\alpha}=\gamma_0\bm{\gamma}$ and $\beta = \gamma_0$ are the Dirac matrices, $M^\ast=M+U_S$ is the effective mass and $u(\bm{p})$ are the Dirac spinors given in Eq.~(\ref{equ17}).
The eigenvalues $E(p)=U_V+E^\ast(p)$ are used for the solution of the Thompson equation (\ref{equ18}) in the next step of the iteration.

Considering the isospin dependence, it is evident that $U_{n}$ and $U_{p}$ in Eq.~(\ref{equ22}) are coupled through the $np$ component of the potential
\begin{eqnarray}%
  U_{n}  &=&U_{nn}+U_{np},
  \label{equ26}\\
  U_{p}  &=&U_{pp}+U_{pn}.
  \label{equ27}
\end{eqnarray}
Therefore they must be solved simultaneously, and the relativistic $G_{\tau_1\tau_2}$-matrix is self-consistently evaluated with the single-particle potentials and the
single-particle energies in the standard RBHF iterative procedure.
Once the solution is converged, the total energy per nucleon in nuclear matter can be calculated by~\cite{Brockmann1990}
\begin{eqnarray}
  \frac{E}{A}&=&\frac{1}{A}\sum_{s_{m},\tau}\int_{0}^{k_{F}^\tau}d^{3}k_{m}%
  \langle m|\bm{\alpha \cdot k}_{m}+\beta M|m\rangle-M%
  \label{equ7}\notag\\
  & &+\frac{1}{2A}\sum_{s_{m},s_{n},\tau_1,\tau_2}
  \int_{0}^{k_{F}^{\tau_1}}d^{3}k_{m}\int_{0}^{k_{F}^{\tau_2}}d^{3}k_{n}%
  \langle mn|G_{\tau_1\tau_2}(W_{\tau_1\tau_2})|mn-nm\rangle.
  \label{equ28}%
  \end{eqnarray}
As mentioned above, we will focus on the calculation of the
potential energy.

\subsection{POTENTIAL ENERGY}

As previously mentioned, the $G$-matrix is directly obtained from the Thompson equation (\ref{equ18}) which is written in the c.~m. frame of the two scattering nucleons.
Thus, Eq.~(\ref{equ28}) should be transformed to the c.~m. frame. This yields for the potential energy, the second line of Eq.~(\ref{equ28}),
\begin{equation}
  \frac{E_V}{A}=\frac{1}{2\rho}\frac{8}{(2\pi)^{3}}
  \sum_{mn}\int^{(k_{F}^{\tau_1}+k_{F}^{\tau_2})/2}d^{3}q
  \int^{\substack{|\bm{q}+\bm{P}|\leqslant k_{F}^{\tau_1}\\
  |\bm{q}-\bm{P}|\leqslant k_{F}^{\tau_2}}}d^{3}P
  \langle\bm{q}mn|G_{{\tau_1}{\tau_2}}(\bm{P},W_{\tau_1\tau_2})
  |\bm{q}mn-nm\rangle,
  \label{equ29}%
\end{equation}
with the total density $\rho=\rho_{n}+\rho_{p}$.
The factor 8 is caused by the transformation from the laboratory frame to the c.~m. frame.
The integral over the c.~m. momentum $\bm{P}$ in Eq.~(\ref{equ29}) can not be performed separately because of the momentum dependence of the $G$-matrix.
Obviously, the angular integrations $\int d\Omega_{P}$ in $\int d^{3}P$ depends not only on the magnitude of the total and the relative momentum but also on their relative direction.

In the literatures one has used the averaged c.~m. momentum approximation \cite{Brueckner1968,Alonso2003} (see Appendix \ref{AppendC}), where the average c.~m. momentum is~defined as
\begin{equation}
  P_{\mbox{av}}^{2}=\frac{\int_{0}^{k_{F}^{n}}d^3k_{1}\int_{0}^{k_{F}^{p}}
  d^3k_{2}P^{2}\delta(q-\frac{1}{2}|\bm{k}_{1}-\bm{k}_{2}|)}{\int_{0}%
  ^{k_{F}^{n}}d^3k_{1}\int_{0}^{k_{F}^{p}}d^3k_{2}\delta(q-\frac{1}%
  {2}|\bm{k}_{1}-\bm{k}_{2}|)}.
  \label{equ30}
\end{equation}
It does not depend on the direction and this value is usually applied in the $G$-matrix in Eq.~(\ref{equ29}).
In this investigation we do not use this approximation
and we focus here on how to carry out the angular integrations $\int d\Omega_{P}=\int\sin\theta d\theta d\phi$ exactly, where $\theta$ is the angle between $\bm{q}$ and $\bm{P}$.
On the basis of the condition $|\bm{q}+ \bm{P}|\leqslant k_{F}^{\tau_1}$ and $|\bm{q}-\bm{P}|\leqslant k_{F}^{\tau_2}$, this leads to restrictions on the angle $\theta$.
Firstly, in order to give a more clear understanding of the calculations in detail, the Fermi sphere method~\cite{Drischler2014} is adopted as a powerful tool to
calculate the angle integral $\int d\Omega_{P}$.
Assuming $k_{F}^{n}\geqslant k_{F}^{p}$, one has to distinguish two cases:
\begin{align}
  \rm{(a)}~~~~&k_{F}^{n}\geqslant3k_{F}^{p}~~~~(\rm{or}~ \alpha\geqslant13/14),
  \label{equ31}\\
  \rm{(b)}~~~~&k_{F}^{n}\leqslant3k_{F}^{p}~~~~(\rm{or}~ \alpha\leqslant13/14).
  \label{equ32}
\end{align}
Moreover, there exist three possible situations depending on the value of $|\bm{q}|$ in both of the cases (a) and (b), and a more complicated problem is that at a given $|\bm{q}|$, there are also several regions depending on the magnitude of $|\bm{P}|$.
The details of all the above formulae are provided in
Appendix \ref{AppendB}.

\begin{figure}[ptbh]
  \centering
  \includegraphics[width=3.5in]{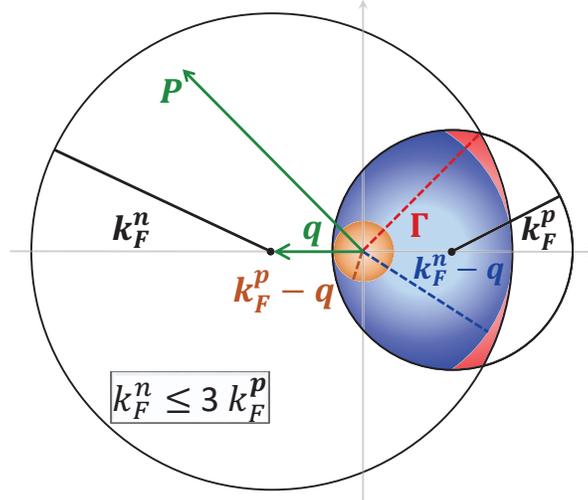} \caption{ (Color online) Different regions contributing to the c.~m. momentum angular integrations in the case of $k_{F}^{n} \leqslant3 k_{F}^{p}$, $k_{F}^{p} \geqslant|\bm{q}| > (k_{F}%
  ^{n}-k_{F}^{p})/2$. Only the overlap (orange, blue and red) contributes to the integral. The orange, blue and red regions denote three angular intervals of integration, which separate the whole space into three parts as given in Eq.~(\ref{equ32}). Green vector-lines represent the c.~m. momentum $\bm{P}$ and
  relative momentum $\bm{q}$. The maximum range of $|\bm{P}|$ for three different parts are denoted by dashed lines. For reference, the Fermi spheres of protons and neutrons are also given with two different Fermi momentum $k_{F}^{n},~k_{F}^{p}$. }%
  \label{angle3}%
\end{figure}

Here we give an example shown in Fig.~\ref{angle3} for the case of $k_{F}^{n} \leqslant3 k_{F}^{p}$, $k_{F}^{p} \geqslant|\bm{q}| >$~$ (k_{F}^{n}-k_{F}^{p})/2$.
Making use of two different Fermi spheres in momentum
space to represent protons and neutrons, displaced by $\pm\bm{q}$ relative to the origin.
Taking the direction of the relative momentum $\bm{q}$
along the horizontal axis, the solid angle $d\Omega_{P}$ is linked with the vector $\bm{P}$.
Considering the integration of the ground state energy in Eq.~(\ref{equ29}), the vector $\bm{P}+\bm{q}=\bm{k_{m}}$ and $\bm{P}-\bm{q}=\bm{k_{n}}$ should be inside of the two solid black Fermi spheres.
This is possible in three different angular intervals
with different colors (orange, blue and red).
The maximum range of $|\bm{P}|$ for the three different parts is represented as $k_{F}^{p}-q$, $k_{F}^{n}-q$ and
$\Gamma= \left[\frac{1}{2}({k_{F}^{n}}^{2}+{k_{F}^{p}}^{2})-
q^{2}\right]^{1/2}$, respectively.
Finally we have,
\begin{equation}
  \int d\Omega_{P}=
  \begin{cases}
  4\pi, & k_{F}^{p}-q \geqslant P \geqslant0\\
  \displaystyle 2\pi\left(  1+\frac{{k_{F}^{p}}^{2}-q^{2}-P^{2}}{2qP} \right)
  , & k_{F}^{n}-q \geqslant P > k_{F}^{p}-q\\
  \displaystyle 2\pi\frac{\frac{1}{2}\left(  {k_{F}^{n}}^{2}+{k_{F}^{p}}%
  ^{2}\right)  -q^{2}-P^{2}}{qP}, & \left[  \frac{1}{2}({k_{F}^{n}}^{2}%
  +{k_{F}^{p}}^{2})-q^{2}\right]  ^{1/2} \geqslant P > k_{F}^{n}-q.
  \end{cases}
  \label{equ33}
\end{equation}



\section{Results and discussion}\label{sec3}

We summarize our results for the properties of nuclear matter in Table \ref{tab1}.
In the first row, we show our RBHF results with (average) and without (exact) the approximation of averaged c.~m. momentum.
The non-relativistic BHF results with and without three-body forces (TBF) are shown in the second row.
For comparison, empirical values are also listed.

In comparison with the results from non-relativistic BHF without three-body forces, the saturation point is shifted towards lower density for relativistic BHF theory using the Bonn potentials. The result for potential Bonn A even meets the empirical region~\cite{Bethe1971,Sprung1972}.
In order to reproduce the saturation point of symmetric matter within non-relativistic BHF, one needs to introduce a three-body force in Ref.~\cite{Vidana2009}.
This three-body force requires two phenomenological parameters that need to be fixed by requiring that the BHF calculation reproduce the energy and saturation density of symmetric nuclear matter.
In Table \ref{tab1}, two sets of such parameters are presented: the original set of Ref.~\cite{Baldo1999} (labeled TBFa), and another new set of Ref.~\cite{Vidana2009} (labeled TBFb), in which the parameter associated with the two pion attractive term has been reduced by 10\%, and the one associated with the phenomenological repulsive term has been increased by 20\% in order to get a smaller saturation density.

\begin{sidewaystable}
\caption{ Bulk parameters (as described in the text) of symmetric and asymmetric nuclear matter at saturation density $\rho_0$. Results obtained in RBHF theory using the Bonn potentials A, B, and C with exact and averaged c.~m. momentum, are compared with those found in non-relativistic BHF theory with and without TBF~\cite{Brockmann1990,ZHLi2006,Vidana2009}. The quantities $\Delta$ are defined as the differences between the exact and the averaged treatment of the c.~m. momentum.
The empirical values are also listed in the last row.
}
\begin{tabular}{c|cccccccccccccccc}
\hline\hline
\multicolumn{2}{c}{\multirow{2}{*}{Model} ~~~~~~~~~~~~~\multirow{2}{*}{Potential}}~~~~~~& & $\rho_0$ & $E/A$ & $K_{\infty}$ & $Q_0$ & $M_0$ & $E_{\mathrm{sym}}$ & $L$ & $K_{\mathrm{sym}}$ & $K_{\mathrm{asy}}$ & $K_{\tau}$ & $K_{\mathrm{Coul}}$ & \\
\multicolumn{2}{c}{} & &  (fm$^{-3}$)~&~(MeV)~&~(MeV)~&~(MeV)~&~(MeV)~&~(MeV)~&~(MeV)~&~(MeV)~&~(MeV)~&~(MeV)~&~ (MeV)~&& \\
\hline
&  & exact~~~~& 0.180 & -15.38 & 286 & 731 & 4163 & 33.7 & 75.8 & -57.0 & -512 & -705 & -8.30 &   \\
& A & average~~~~& 0.182 & -15.04 & 289 & 650 & 4118 & 32.6 & 74.7 & -53.1 & -501 & -669 & -8.09 &  \\
&  &$\Delta$~~~~& -0.002 & -0.34 & -3 &  81 & 45 & 1.1 &  1.1 &  -3.9 & -11 & -36 & -0.21 &  \\
\cline{2-15}
&  & exact~~~~& 0.164 & -13.44 & 222 & 547 & 3211 & 29.9 & 63.0 & -56.3 & -434 & -590 & -7.98 &   \\
RBHF~~~~~ & B & average~~~~& 0.165 & -13.08 & 220 & 791 & 3431 & 28.7 & 65.3 &
-47.5 & -439 & -674 & -8.86 & \\
&  &$\Delta$~~~~& -0.001 & -0.36 &  2 & -244 & -220 & 1.2 & -2.3 &  -8.8 & 5 & 84  &  0.88 &  \\
\cline{2-15}
&  & exact~~~~& 0.149 & -12.12 & 176 & 260 & 2372 & 26.8 & 51.7 & -55.6 & -366 & -442 & -7.00 &   \\
& C & average~~~~& 0.150 & -11.75 & 168 & 638 & 2654 & 25.6 & 58.8 &
-41.1 & -394 & -618 & -8.74 &\\
&  &$\Delta$~~~~& -0.001 & -0.37 &  8 & -378 & -282 & 1.2 & -7.1 & -14.5 & 28 & 176 &  1.74 &  \\
\hline
& A  & & 0.428 & -23.55 & 204 &  & &32.1 & & & & & &\\
\multirow{4}{*}{BHF}~~~~~ & B  & & 0.309 & -18.30 & 160 &  & &31.8 &\\
& C  & & 0.247 & -15.75 & 143 &  & &28.5 &\\
&AV18& W/O TBF~~~~& 0.240 & -17.30 & 214 & -225 & 2343 & 35.8 & 63.1 & -27.8 & -406 & -340 & -6.01  \\
&AV18& TBFa~~~~& 0.187 & -15.23 & 196 & -281 & 2071 & 34.3 & 66.5 & -31.3 & -430 & -335 & -5.23  \\
&AV18& TBFb~~~~& 0.176 & -14.62 & 186 & -225 & 2007 & 33.6 & 66.9 & -23.4 & -425 & -344 & -5.30  \\
\hline
\multicolumn{2}{c}{\multirow{2}{*}{Empirical}}~~~~~~~~~~~~~~~~~~~~~~~~~~~ & & 0.166 & -16 & 240 &  & & 32 & 88 &  &  & -550 & \\
\multicolumn{2}{c}{}     & & $\pm$0.018 & $\pm$1 & $\pm$20 & & & $\pm$ 2 & $\pm$25 &  &  & $\pm$100 & \\
\hline\hline
\end{tabular}
\label{tab1}
\end{sidewaystable}

In the sixth column of Table \ref{tab1} we show the symmetry energy at saturation density. For the exact calculation it turns out to be 33.7 MeV, which is in good agreement with the empirical value of 32 $\pm$ 2 MeV~\cite{STEINER2005}.

Using the previously RBHF method, the incompressibility of nuclear matter at saturation density is 286 MeV for the potential Bonn A and about 222 MeV for the potential Bonn B, which is in satisfactory agreement with the commonly accepted value of 240~$\pm$~20 MeV~\cite{Lietal2007,Garg2007,GargColo2018}.
It should be noted that, after including the three-body forces within non-relativistic BHF, the incompressibility coefficient decreases considerably and reaches values far from the lower bound of $K_{\infty}=220$ MeV imposed by experiments.

At present, there is no experimental constraint on $Q_0$, which is defined as the third density derivative of the symmetric nuclear matter energy at saturation.
The microscopic predictions of RBHF theory for $Q_0$ are large and positive.
They are in contrast to the non-relativistic BHF results with negative values.
As a consequence, in Eq.~(\ref{equ14}) the values of
$K_{\mathrm{Coul}}=\frac{3}{5}\frac{e^{2}}{r_{0}}\left(  -8-\frac{Q_{0}}{K_{\infty}}\right)$ are larger for RBHF than those found in non-relativistic BHF.
We also see that the approximate expression (\ref{equ16})
$K_{\tau}\approx K_{\mathrm{asy}}=K_{\mathrm{sym}}-6L$%
~\cite{ChenKoLi2005,LiChenKo2008,Centelles2009}, which
is quite often used instead of
$K_{\tau}=K_{\mathrm{asy}}-(Q_{0}/K_{\infty})L$,
can lead to a remarkable difference in $K_{\tau}$.
The results of this addendum indicate that generally the higher order $Q_{0}$ contribution to $K_{\tau}$ can not be neglected, neither in relativistic nor in non-relativistic BHF, especially for larger $L$ values.

It is shown that the saturation densities do not change substantially for the exact treatment of the c.~m. momentum as compared to the results of the averaged c.~m. momentum approximation.
It is a common characteristic of the results for three different nucleon-nucleon potentials (Bonn A, B, and C), that the exact treatment of the c.~m. momentum produces small, but non-negligible contributions to the binding energy per nucleon at saturation densities, compared with the results of the conventional averaged c.~m. momentum
approximation.
These non-negligible differences in the binding energy are
important, when studying effects of higher order physical quantities in both of the energy in symmetric nuclear matter and the symmetry energy.
For some of the properties associated with the EoS, such as $\rho_{0}$, $E/A$, $K_{\infty}$, $E_{\mathrm{sym}}$ and $L$, the differences are relatively small, but they become significant for the remaining higher order parameters.
Especially we find significant differences for the quantities $Q_{0}$, $M_{0}$, $K_{\mathrm{sym}}$, $K_{\mathrm{asy}}$, $K_{\mathrm{Coul}}$, and $K_{\tau}$.

\begin{figure}[!htbp]
  \centerline{
  \includegraphics[width=8.0cm]{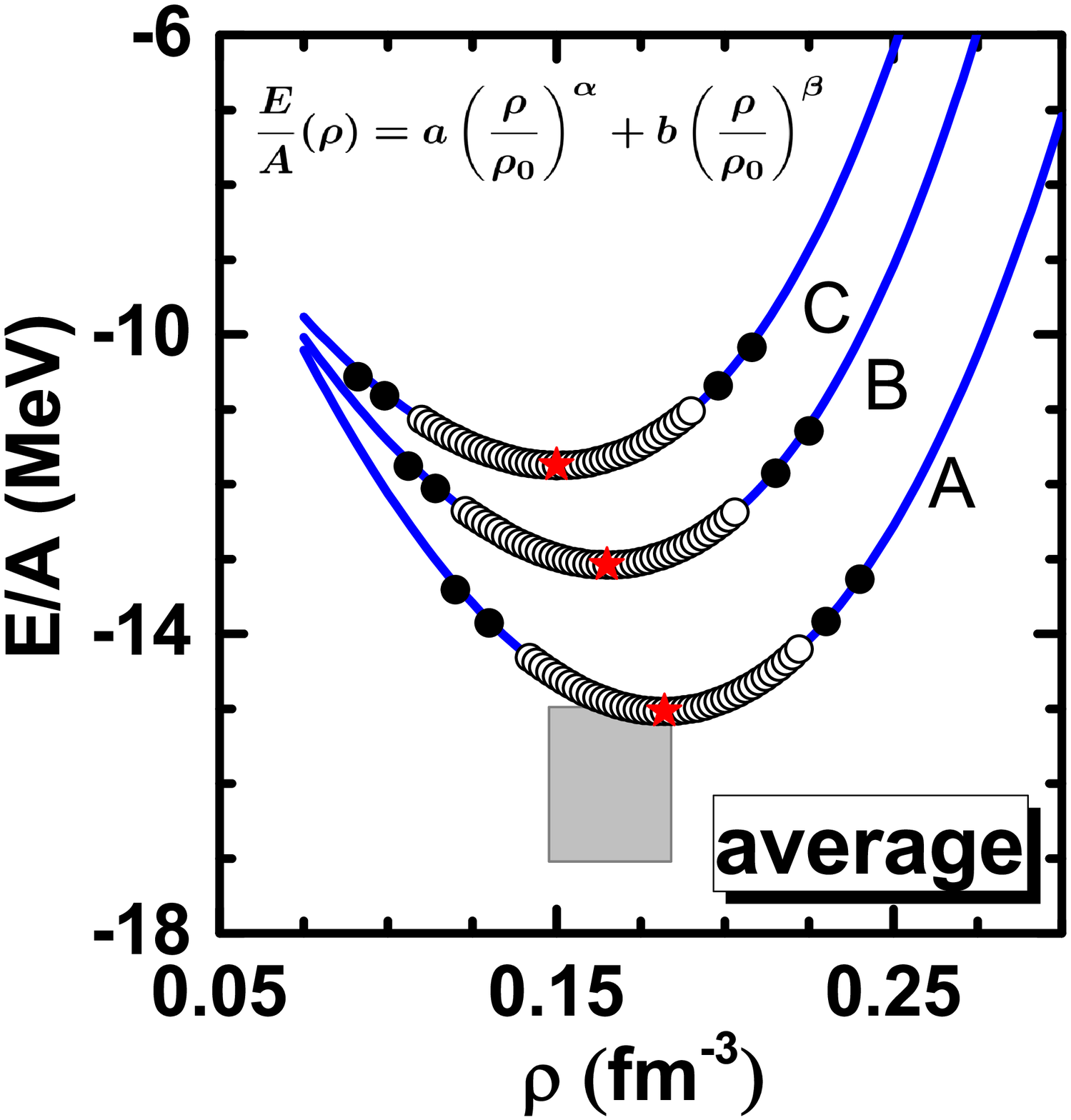}\includegraphics[width=8.08cm]{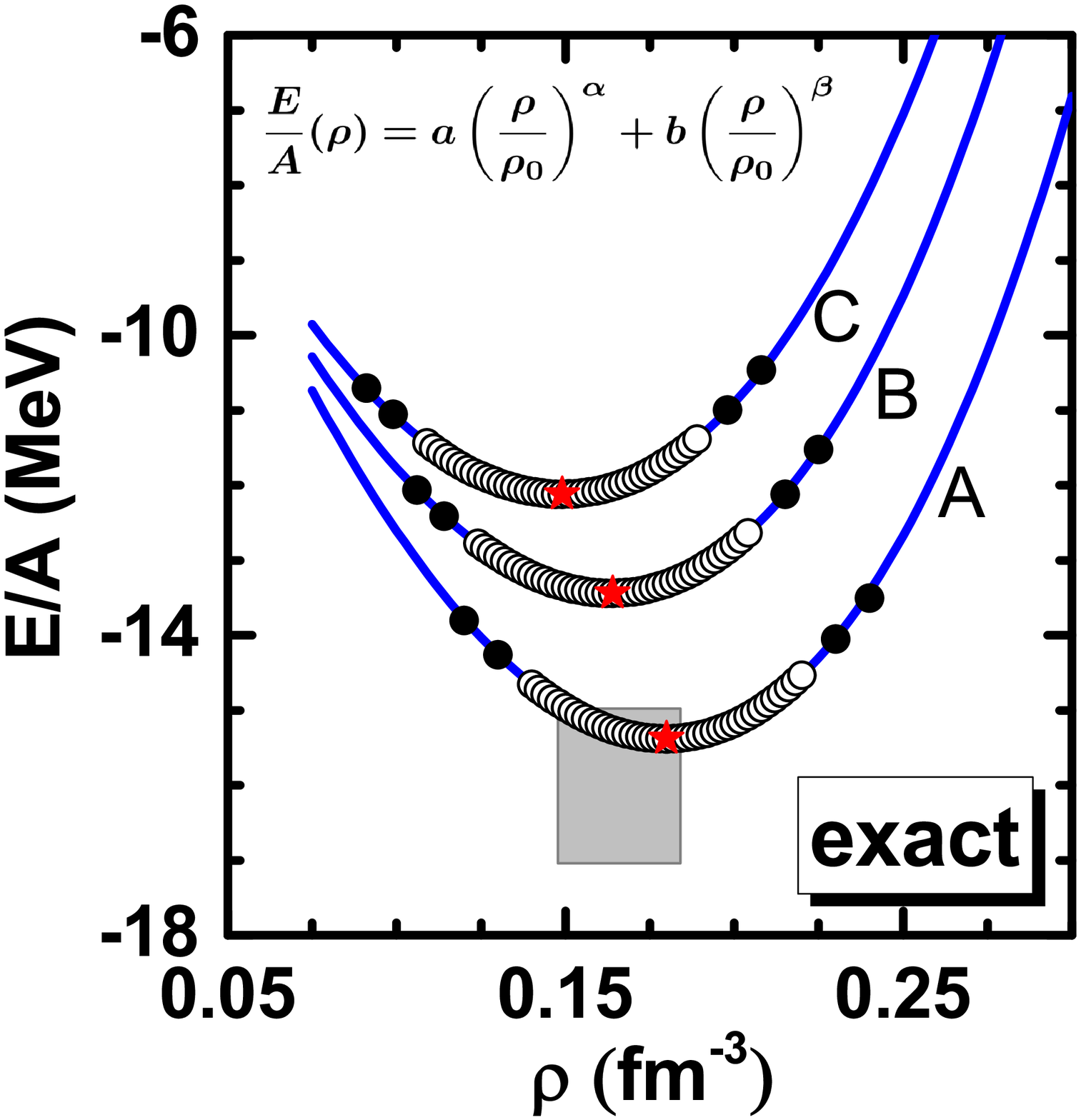}
  } \caption{ (Color online) Binding energy per nucleon for nuclear matter as a function of the total density $\rho$. Results for Bonn potentials A, B, C with (left panel) and without (right panel) c.~m. momentum approximation are shown.
  The RBHF results are represented by open and solid circles, where open circles stand for the data used in the fit and solid circles for examining the validity of the results of the fit (solid curves). The red stars indicate the saturation points obtained from RBHF results.}%
  \label{Graph6}%
\end{figure}

In order to have accurate expressions for the various quantities defined as the density derivatives of the energy of symmetric nuclear matter and of the symmetry energy in Table \ref{tab1}, we parameterized the energy of symmetric nuclear matter and the symmetry energy in vicinity of the saturation density $\rho_{0}$.
It has been found that the EoS can be accurately represented using the following functional form~\cite{Gandolfi2012}:
\begin{equation}
  \frac{E}{A}(\rho)=a\left(  \frac{\rho}{\rho_{0}}\right)  ^{\alpha}+b\left( \frac{\rho}{\rho_{0}}\right) ^{\beta},
  \label{equ34}
\end{equation}
where $E/A$ is the binding energy per nucleon as a function of the nuclear density $\rho$, and the parameters $a$, $\alpha$, $b$, and $\beta$ are obtained by fitting the RBHF theory using the Bonn potentials.
In a similar way, a two-parameter representation for the symmetry energy around saturation density is frequently used~\cite{LiChenKo2008}:
\begin{equation}
  E_{\mathrm{sym}}(\rho)=c\left(  \frac{\rho}{\rho_{0}}\right)  ^{\gamma}.
  \label{equ35}%
\end{equation}
The results of these fits, i.e. the parametrization of the equations of state obtained with and without c.~m. momentum approximation are shown in Table \ref{tab2} and in Fig.~\ref{Graph6}.
As we can see, the binding energy calculated by RBHF theory using the potential Bonn A without c.~m. momentum approximation agrees better with the empirical value than the results based on the c.~m. momentum approximation.

\begin{table}[htbp]
\caption{Fit parameters for the nuclear matter properties defined in Eqs.~(\ref{equ34}) and (\ref{equ35}) for RBHF theory using the Bonn potentials A, B, and C. }%
\begin{tabular}
[c]{c|ccccccccccccc}\hline\hline
\multicolumn{2}{c}{\multirow{2}{*}{Model} ~~~~~\multirow{2}{*}{Potential}} &
~~~ & $a$ & $\alpha$ & $b$ & $\beta$ & $c$ & $\gamma$ &  &  &  & \\
\multicolumn{2}{c}{} &  & (MeV) &  & (MeV) &  & (MeV) &  &  &  &  &  &
\\\hline
& \multirow{2}{*}{A} & exact~~ & -19.25 & 0.64 & 3.87 & 3.21 & 33.72 & 0.75 &
&  &  &  & \\
&  & average~~ & -19.53 & 0.70 & 4.49 & 3.05 & 32.63 & 0.76 &  &  &  &  &
\\\cline{2-11}%
\multirow{2}{*}{RBHF}~~~ & \multirow{2}{*}{B} & exact~~ & -16.23 & 0.56 &
2.79 & 3.26 & 29.92 & 0.70 &  &  &  &  & \\
&  & average~~ & -15.15 & 0.51 & 2.07 & 3.69 & 28.73 & 0.76 &  &  &  &  &
\\\cline{2-11}
& \multirow{2}{*}{C} & exact~~ & -14.89 & 0.55 & 2.77 & 2.94 & 26.85 & 0.64 &
&  &  &  & \\
&  & average~~ & -13.15 & 0.41 & 1.40 & 3.86 & 25.57 & 0.77 &  &  &  &  &
\\\hline\hline
\end{tabular}
\label{tab2}
\end{table}

\begin{figure}[tbh]
  \centerline{
  \includegraphics[width=8.0cm]{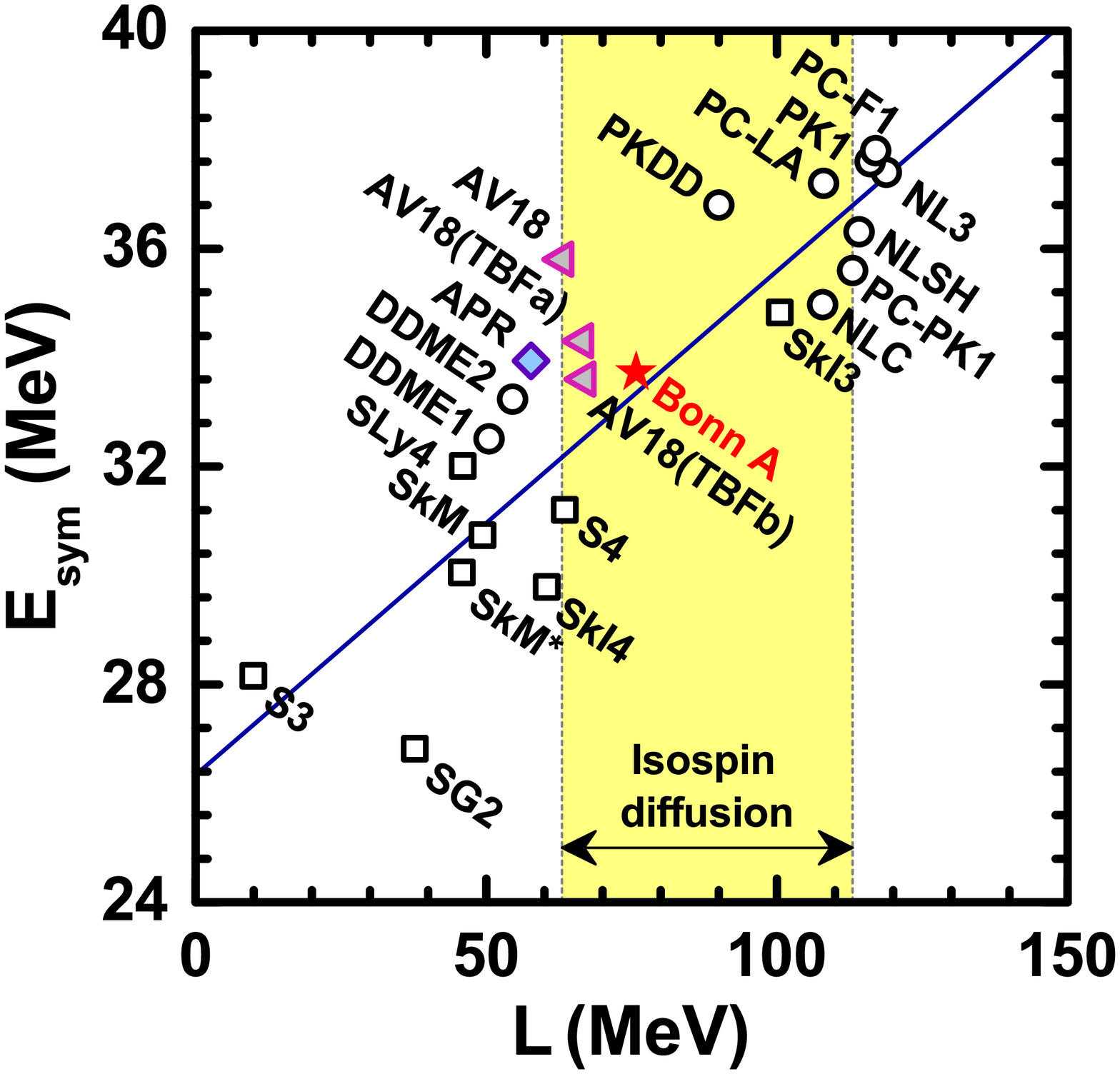}
  \includegraphics[width=8.43cm]{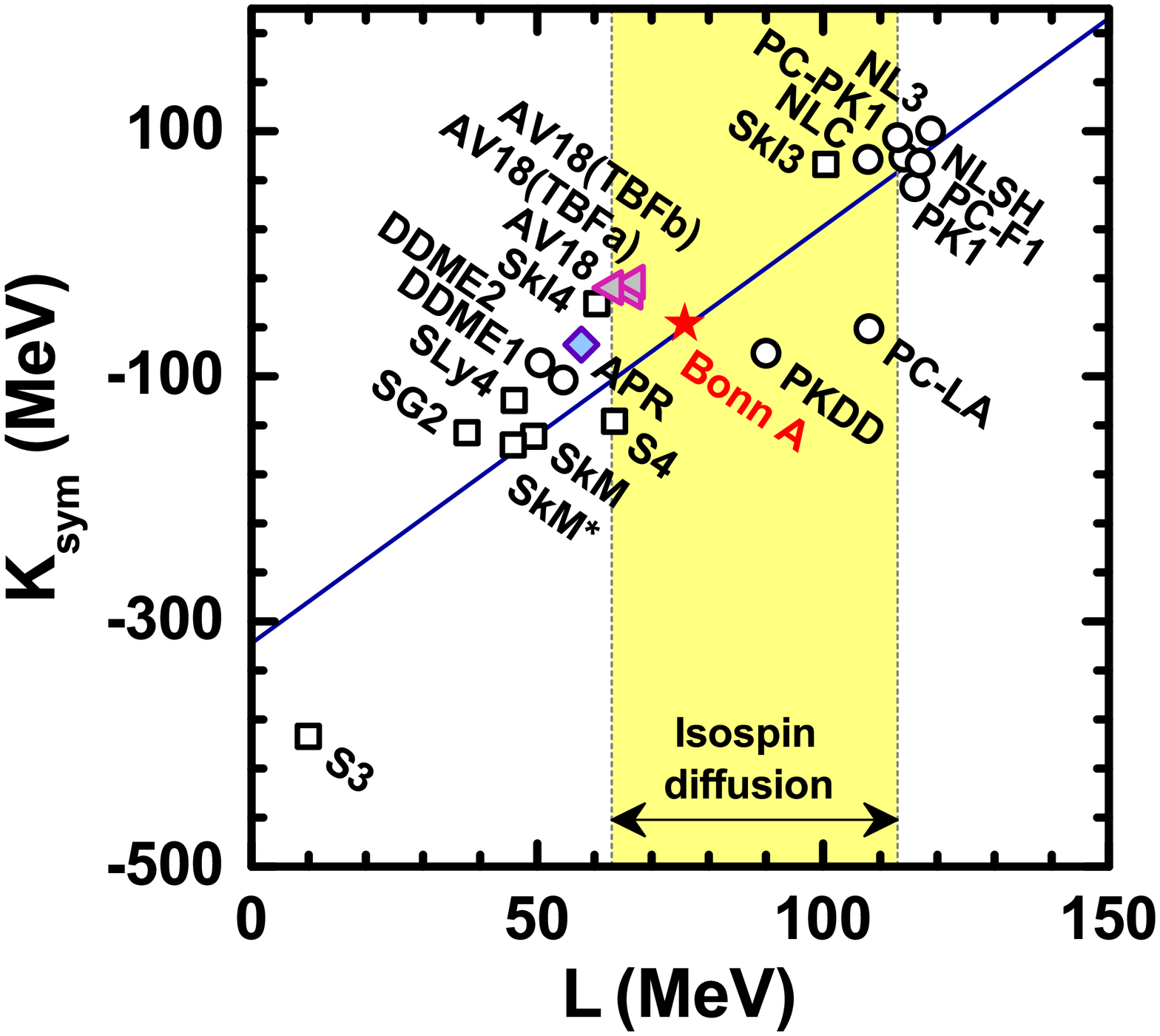}
  }
  \caption{ (Color online) $E_{\mathrm{sym}}$ (left panel) and
  $K_{\mathrm{sym}}$ (right panel) with $L$ calculated by RBHF theory using the potential Bonn A (red star), in comparison with results obtained by BHF (triangles)~\cite{Vidana2009}, variational methods APR (diamond)~\cite{A.AkmalRavenhall1998} and various density functionals (circles and squares)~\cite{ZhaoLiYaoEtAl2010,H.SagawaZhang2007}. The shaded regions denote
  the constraints on $L$ from isospin diffusion
  ~\cite{LiChenKo2008,ChenKoLi2005,ChenKoLi2005a}. The blue line is the linear fit to
  the results of density functionals.}%
\label{Graph1}%
\end{figure}

In Fig.~\ref{Graph1} we show the correlations between $L$ and $E_{\mathrm{sym}}$ (left panel) and between $L$ and $K_{\mathrm{sym}}$ (right panel), which have been investigated in Ref.~\cite{ChenKoLi2005a,Vidana2009}.
The values of $E_{\mathrm{sym}}$ and $K_{\mathrm{sym}}$ obtained from both the non-relativistic (squares) and relativistic (circles) density functionals exhibit a linear correlation with $L$.
It should be mentioned that the result of RBHF theory using the potential Bonn A is in excellent agreement with this tight correlation.
In addition, other \textit{ab-initio} calculations, such as the results of non-relativistic BHF and the variational microscopic calculations of Akmal, Pandharipande and Ravenhall (hereafter APR) which incorporate relativistic boost corrections and three-nucleon interactions~(using the A18+$\delta v$+UIX$^{\ast}$ interaction)~\cite{A.AkmalRavenhall1998} are also given.
It can be seen that these two correlations also exist in
microscopic approaches.
Note that the RBHF results for $L$ are also located
inside the region constrained by the isospin diffusion data~\cite{LiChenKo2008,ChenKoLi2005,ChenKoLi2005a}.
According to Fig.~\ref{Graph1}, it is clear that the symmetry energy $E_{\mathrm{sym}}$ and the curvature parameter~$K_{\mathrm{sym}}$ are both sensitive to the slope parameter $L$, increasing almost linearly with
increasing~$L$.
There is no direct experimental information on the
$K_{\mathrm{sym}}$ parameter.
However, as proposed in Ref.~\cite{ChenCaiKoEtAl2009}, once accurate experimental information becomes available for $L$, these correlations could be exploited to obtain theoretical
estimates for $K_{\mathrm{sym}}$.

\begin{figure}[tbh]
  \centerline{
  \includegraphics[width=9.0cm]{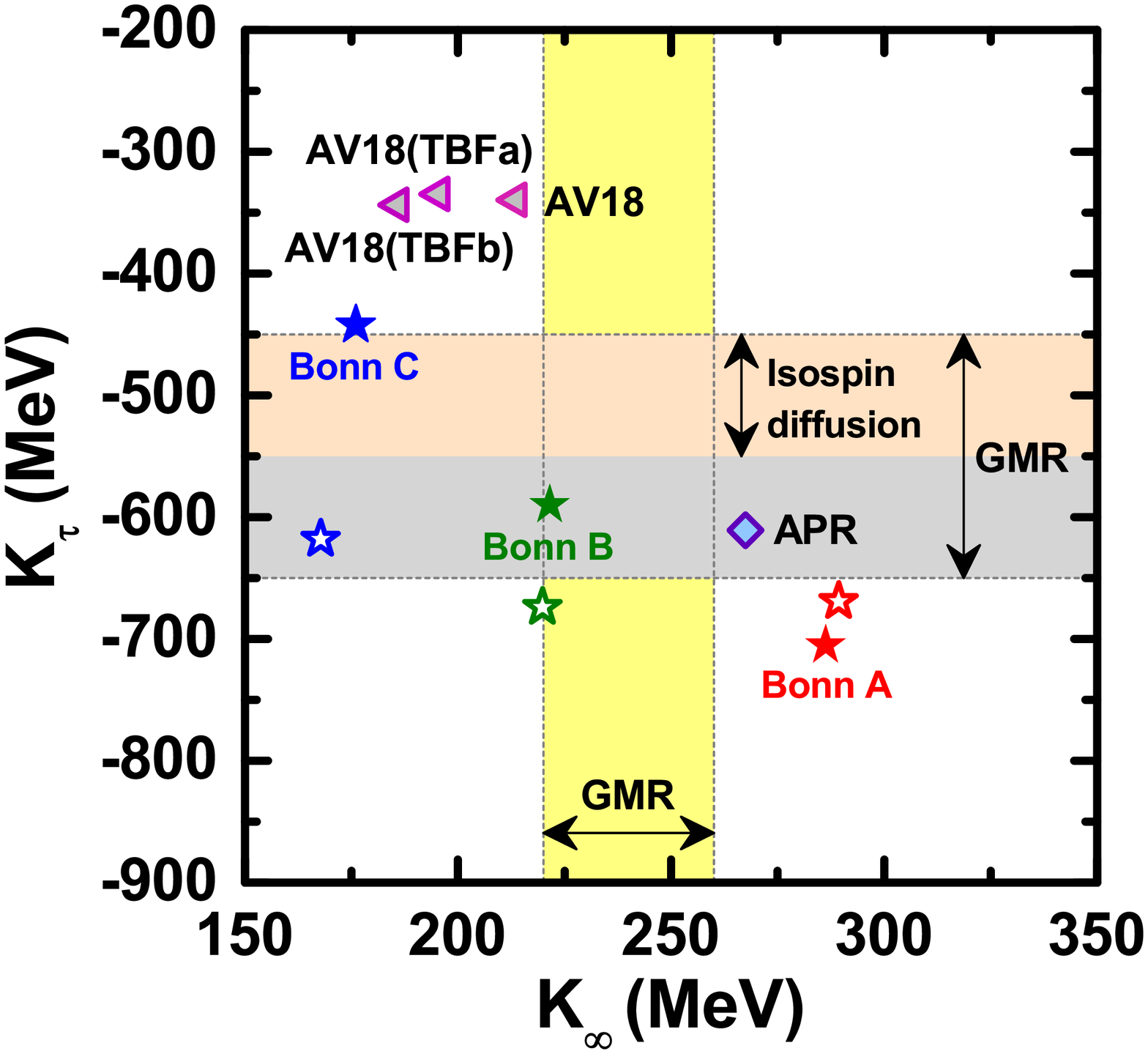}\includegraphics[width=9.0cm]{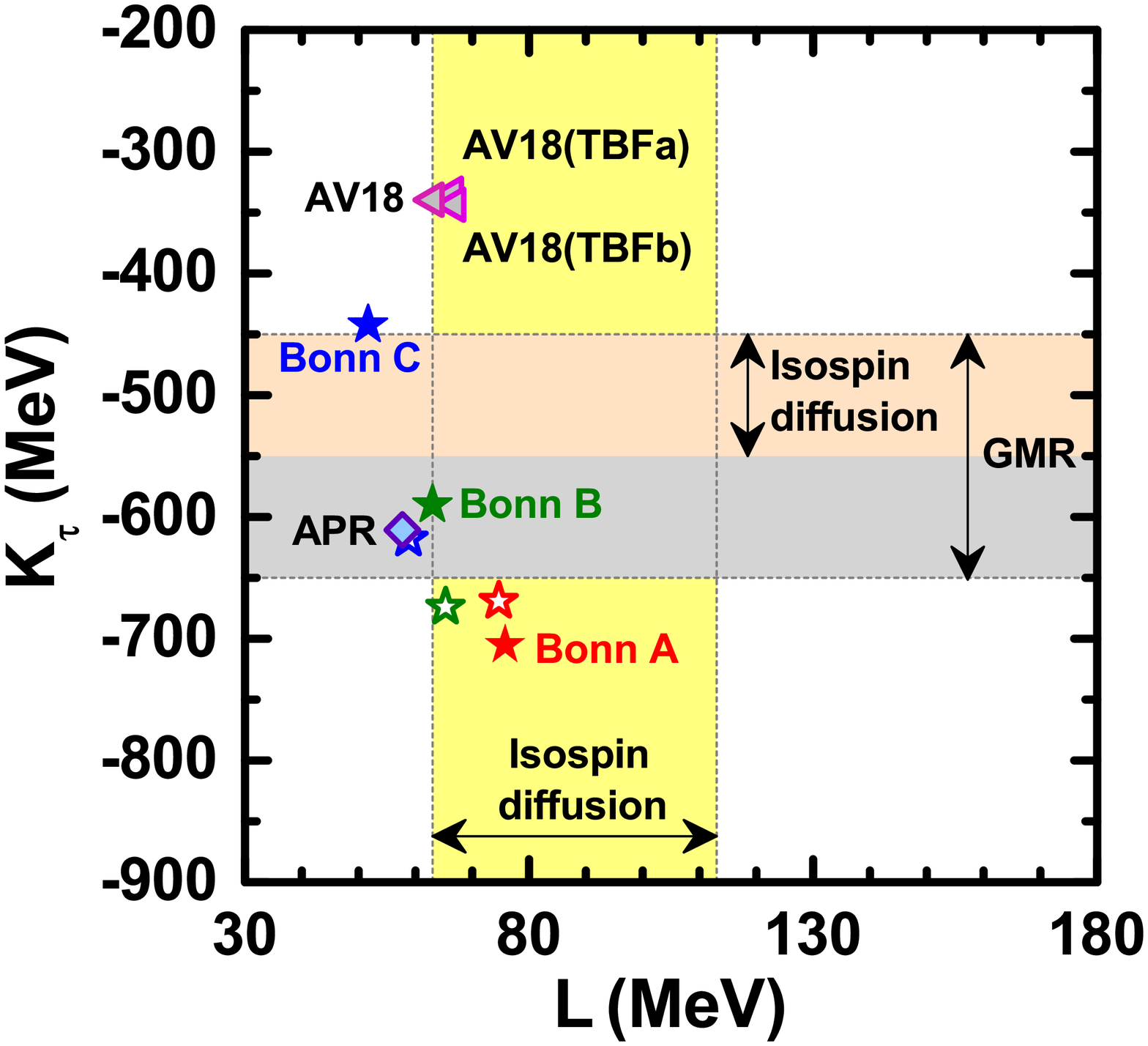}
  }
  \caption{ (Color online) Values of $K_{\infty}$ (left panel) and $L$ (right panel) with $K_{\tau}$ calculated by RBHF theory using the Bonn potentials with (open star) and without (solid star) averaged c.m. momentum approximation, compared with BHF (triangles)~\cite{Vidana2009} and APR
  (diamond)~\cite{A.AkmalRavenhall1998}.
  The shaded regions indicate the experimental ranges of $K_{\tau}$ and $K_{\infty}$ from the GMR of Sn isotopes
  ~\cite{Lietal2007,GargColo2018} and on $K_{\tau}$ and $L$ as determined in~\cite{LiChenKo2008,ChenKoLi2005,ChenKoLi2005a} from isospin diffusion. }%
  \label{Graph3}%
\end{figure}

One can see in Fig.~\ref{Graph3} the values of $K_\infty$, $L$, and $K_\tau$ for the present RBHF calculations using the Bonn potentials and compare with the predictions of BHF, and APR as given in Table \ref{tab1}. The shaded rectangular regions encompass the recent values of $K_\infty=240\pm20$~MeV~\cite{Lietal2007,GargColo2018}, $K_\tau=-550\pm 100$~MeV~\cite{Lietal2007,GargColo2018} and $L=88\pm25$~MeV~\cite{LiChenKo2008,ChenKoLi2005,ChenKoLi2005a}.
The experimental values obtained from the GMR and from isospin diffusion for $K_\tau$, $K_\infty$, and $L$ together provide a way to choose the most appropriate interaction used in the EoS calculations.
Although a majority of the interactions fail to meet this region established by these measurements, it is worth mentioning here that the RBHF theory using the potential Bonn B without c.~m. momentum approximation is within this region. It has been shown by Sagawa et al.~\cite{H.SagawaZhang2007}, that $K_\tau$ is largely negative and shows an anti-correlation with the nuclear matter incompressibility $K_\infty$ in both of the non-relativistic and relativistic density functionals, that is, any approach that has a larger $K_\infty$ gives a smaller $K_\tau$. The same conclusions have been verified in the microscopic calculations.

\begin{figure}[!htbp]
  \centerline{
  \includegraphics[width=9.0cm]{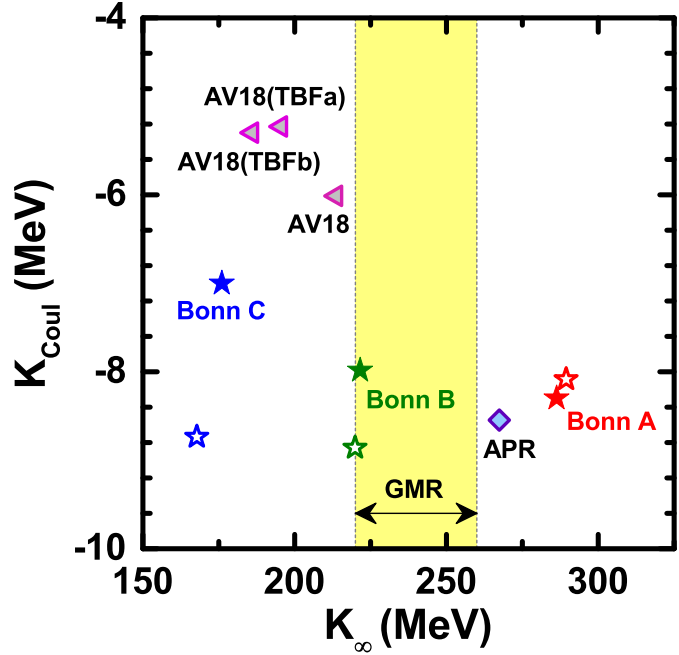}
  }
  \caption{
    (Color online) Values of $K_\mathrm{Coul}$ and $K_\mathrm{\infty}$ calculated by RBHF theory using the Bonn potentials with (open star) and without (solid star) averaged c.~m. momentum approximation, compared with BHF (triangles)~\cite{Vidana2009} and APR (diamond)~\cite{A.AkmalRavenhall1998}. The shaded regions denote the constraints on $K_\mathrm{\infty}$ from the GMR of Sn isotopes~\cite{Lietal2007,GargColo2018}.
  }
  \label{Graph5}
\end{figure}

As noted previously in Eq.~\eqref{equ12}, the incompressibility $K_A$ of finite nuclei may be parameterized as~\cite{Blaizot1980}:
\begin{equation}
  K_A \approx K_{\infty}+K_{\mathrm{surf}}A^{-1/3}+K_\tau\alpha^2+K_{\mathrm{Coul}}
  \frac{Z^2}{A^{4/3}}.
  \label{equ36}
\end{equation}
$K_{\mathrm{Coul}}$ is essentially a model-independent
term (in the sense that the deviations from one theoretical
model to another are quite small)~\cite{H.SagawaZhang2007}. Therefore, in order to obtain $K_\tau$, an approximately quadratic relation between $K_A-K_{\mathrm{Coul}}Z^2A^{-4/3}$ and the asymmetry parameter $\alpha$ can be used to fit the experimental data. In Refs.~\cite{Lietal2007,Pateletal2012}, a value of $-5.2~\pm~0.7$ MeV has been applied for $K_{\mathrm{Coul}}$ which has been derived  from 13 parameter sets of the Skyrme interaction~\cite{H.SagawaZhang2007}, and the uncertainty in the value of $K_{\mathrm{Coul}}$ contributes $\sim$ 15 MeV to $K_\tau$ from the measurement of the GMR in even-A Sn isotopes and $\sim$ 20 MeV in even-A Cd isotopes.

As discussed earlier, we can see the values of $K_{\mathrm{Coul}}$ derived from RBHF theory using the Bonn potentials are larger than those derived from BHF because the skewness parameters $Q_0$ for RBHF theory are large and positive. In addition, the values of $K_{\mathrm{Coul}}$ provided by the relativistic approaches including RBHF are larger than $-5.2~\pm~0.7$ MeV, which indicates that higher order corrections~(e.g. $Q_0$)~play an important role in $K_{\mathrm{Coul}}$. Therefore it is important to study the effects on $K_\tau$ derived from relativistic approaches when using different values of $K_{\mathrm{Coul}}$.

\section{Summary}\label{sec4}

Quantities like the binding energy of symmetric nuclear matter and the symmetry energy and their density dependence
play an important role in modern nuclear physics and astrophysics. Non-relativistic and relativistic Brueckner-Hartree-Fock theory allows an \textit{ab-initio} derivation of these quantities from the experimentally known bare nucleon-nucleon interaction. In the present paper, we derived an exact and analytic expression of the angular integrations for the c.~m. momentum $\bm{P}$ by employing the Fermi sphere method, which is important for a precise numerical calculation of the binding energy, especially for asymmetric nuclear matter.
In order to examine the effect of the exact treatment of the c.~m. momentum  and to assess the reliability of the averaged c.~m. momentum approximation, we have systematically studied the density dependence of the energy of symmetric nuclear matter and of the symmetry energy in vicinity of the saturation density $\rho_0$, within relativistic Brueckner-Hartree-Fock theory using the Bonn potentials with and without averaged c.~m. momentum approximation.

Our results clarified that for some of the properties, such as $\rho_0$, $E/A$, $K_{\infty}$, $E_{\mathrm{sym}}$ and $L$, the approximation of an averaged c.~m. momentum is quantitatively reliable, but for the remaining higher order parameters, such as $Q_{0}$, $M_{0}$, $K_{\mathrm{sym}}$, $K_{\mathrm{asy}}$, $K_{\mathrm{Coul}}$ and $K_{\tau}$ there are considerable discrepancies between the exact treatment of the angle integrations and the angle-averaged approximation.

Furthermore, the results of our relativistic calculations have been compared with those of non-relativistic BHF theory. It turns out that the saturation density $\rho_0$, the binding energy per particle $E/A$, and the incompressibility $K_{\infty}$ derived from RBHF theory agree better with the empirical values than those from non-relativistic BHF theory.

We have also studied the correlation between the $L$ and $E_{\mathrm{sym}}$ and $L$ and $K_{\mathrm{sym}}$. It is found that the results of RBHF, BHF and variational calculations (APR) are in excellent agreement with the tight correlations already obtained by other calculations using non-relativistic and relativistic density functionals. This agreement suggests that these correlations are not only due to the mean field nature of these approaches but also exist in the microscopic methods. We have confirmed for the microscopic methods that there is an anti-correlation between the symmetry term $K_\tau$ and the incompressibility $K_{\infty}$, a trend pointed out by Sagawa et al. ~\cite{H.SagawaZhang2007}.
In addition, we note that the microscopic predictions for $Q_0$ from RBHF theory are large and positive, which are in contrast to the non-relativistic BHF theory with negative values.
Our results indicate that generally the higher order $Q_0$ contribution to $K_\tau$ cannot be neglected, and that the value of the higher order corrections~$Q_0$ play an important role for $K_{\mathrm{Coul}}$.

\begin{acknowledgments}
This work was partly supported by the National Key R\&D Program of China (2018YFA0404400), the National Natural Science Foundation of China (NSFC) under Grants No. 11335002, No. 11621131001 and Grants No. 11775099, the China Postdoctoral Science Foundation under Grants No.~2016M600845, No.~2017T100008, and the DFG (Germany) cluster of excellence \textquotedblleft Origin and Structure of Universe\textquotedblright\ (www.universe-cluster.de).
\end{acknowledgments}

\begin{appendix}
\section{Angle-averaged Pauli operator}\label{AppendA}

The definition of the angle-averaged Pauli operator is
\begin{equation}\label{equA1}
  Q_{\tau_1\tau_2}^{\mbox{av}}(k,P)=\frac{\int Q_{\tau_1\tau_2}(\bm{k},\bm{P}) d\Omega}{\int d\Omega},
\end{equation}
where $\Omega$ is the angle between $\bm{k}$ and $\bm{P}$.
We have to distinguish two cases, depending on the values of $k$, $P$, $k_F^p$, $k_F^n$:

(a).~~$(k_{F}^n-k_{F}^p)/2 \geqslant P \geqslant 0 $
\begin{equation}
  Q_{\tau_1\tau_2}^{\mbox{av}}(k,P)=
  \begin{cases}
    0 ,& k< k_{F}^n-P , \\
    \displaystyle \frac{1}{2}\left(\frac{k^2+P^2-{k_{F}^n}^2}{2Pk}+1\right) ,& {k_{F}^n}-P \leq k < {k_{F}^n}+P,\\
    1,& {k_{F}^n}+P \leq k.
  \end{cases}
\label{equA2}
\end{equation}

(b).~~$(k_{F}^n+k_{F}^p)/2 \geqslant P > (k_{F}^n-k_{F}^p)/2 $
\begin{equation}
      Q_{\tau_1\tau_2}^{\mbox{av}}(k,P)=
      \begin{cases}
       0 ,& \displaystyle  k<\left[\frac{1}{2}\left({k_{F}^n}^2+{k_{F}^p}^2\right)-P^2\right]^{1/2}, \\
      \displaystyle \frac{P^2+k^2-\frac{1}{2}[(k_{F}^n)^2+(k_{F}^p)^2]}{2Pk} ,&
      \displaystyle \left[\frac{1}{2}\left({k_{F}^n}^2+{k_{F}^p}^2\right)-P^2\right]^{1/2} \leq k < {k_{F}^p}+P, \\
      \displaystyle  \frac{1}{2}\left(\frac{k^2+P^2-{k_{F}^n}^2}{2Pk}+1\right) ,&  {k_{F}^p}+P \leq k < {k_{F}^n}+P,\\
      1,& {k_{F}^n}+P \leq k.
    \end{cases}
\label{equA3}
\end{equation}

\section{Exact angular integrations for the c.~m. momentum}\label{AppendB}

In this case we have the following possibilities:

\subsection{$k_F^n \geqslant 3k_F^p$ }

(a).~~$ k_{F}^p \geqslant q \geqslant 0 $
\begin{equation}
  \int d\Omega_P=
  \begin{cases}
     4\pi,  &k_{F}^p-q \geqslant P \geqslant 0 \\
    \displaystyle 2\pi\left(1+\frac{{k_{F}^p}^2-q^2-P^2}{2qP} \right), &k_{F}^p+q \geqslant P > k_{F}^p-q.
  \end{cases}
\label{equB1}
\end{equation}

(b).~~$ (k_{F}^n-k_{F}^p)/2 \geqslant q > k_{F}^p $
\begin{equation}
  \int d\Omega_P=
    \displaystyle 2\pi\left(1+\frac{{k_{F}^p}^2-q^2-P^2}{2qP} \right),~~q+k_{F}^p \geqslant P \geqslant q-k_{F}^p.
\label{equB2}
\end{equation}

(c).~~$ (k_{F}^n+k_{F}^p)/2 \geqslant q >  (k_{F}^n-k_{F}^p)/2 $
\begin{equation}
  \int d\Omega_P=
  \begin{cases}
    \displaystyle 2\pi\left(1+\frac{{k_{F}^p}^2-q^2-P^2}{2qP} \right),&k_{F}^n-q \geqslant P \geqslant q-k_{F}^p\\
     \displaystyle 2\pi \frac{\frac{1}{2}\left({k_{F}^n}^2+{k_{F}^p}^2\right)-q^2-P^2}{qP},&
    \left[\frac{1}{2}({k_{F}^n}^2+{k_{F}^p}^2)-q^2\right]^{1/2} \geqslant P > k_{F}^n-q.
  \end{cases}
  \label{equB3}
\end{equation}

\subsection{$k_F^n \leqslant 3k_F^p$ }

(a).~~$(k_{F}^n-k_{F}^p)/2 \geqslant q \geqslant 0 $
\begin{equation}
  \int d\Omega_P=
  \begin{cases}
     4\pi,  &k_{F}^p-q \geqslant P \geqslant 0 \\
    \displaystyle 2\pi\left(1+\frac{{k_{F}^p}^2-q^2-P^2}{2qP} \right), &k_{F}^p+q \geqslant P > k_{F}^p-q.
  \end{cases}
\label{equB4}
\end{equation}

(b).~~$ k_{F}^p \geqslant q > (k_{F}^n-k_{F}^p)/2 $
\begin{equation}
  \int d\Omega_P=
  \begin{cases}
     4\pi, &k_{F}^p-q \geqslant P \geqslant 0 \\
    \displaystyle 2\pi\left(1+\frac{{k_{F}^p}^2-q^2-P^2}{2qP} \right),&k_{F}^n-q \geqslant P > k_{F}^p-q\\
     \displaystyle 2\pi \frac{\frac{1}{2}\left({k_{F}^n}^2+{k_{F}^p}^2\right)-q^2-P^2}{qP},&
    \left[\frac{1}{2}({k_{F}^n}^2+{k_{F}^p}^2)-q^2\right]^{1/2} \geqslant P > k_{F}^n-q.
  \end{cases}
\label{equB5}
\end{equation}

(c).~~$ (k_{F}^n+k_{F}^p)/2 \geqslant q >  k_{F}^p $
\begin{equation}
  \int d\Omega_P=
  \begin{cases}
    \displaystyle 2\pi\left(1+\frac{{k_{F}^p}^2-q^2-P^2}{2qP} \right),&k_{F}^n-q \geqslant P \geqslant q-k_{F}^p\\
     \displaystyle 2\pi \frac{\frac{1}{2}\left({k_{F}^n}^2+{k_{F}^p}^2\right)-q^2-P^2}{qP},&
    \left[\frac{1}{2}({k_{F}^n}^2+{k_{F}^p}^2)-q^2\right]^{1/2} \geqslant P > k_{F}^n-q.
  \end{cases}
\label{equB6}
\end{equation}

\section{Averaged center of mass momentum}\label{AppendC}

The definition of the average c.~m. momentum is~\cite{Brueckner1968}
\begin{equation}
  P_{\mbox{av}}^2=
  \frac{\int_0^{k_F^n}d^3k_1\int_0^{k_F^p}d^3k_2P^2\delta(q-\frac{1}{2}|\bm{k}_1-\bm{k}_2|)}
  {\int_0^{k_F^n}d^3k_1\int_0^{k_F^p}d^3k_2\delta(q-\frac{1}{2}|\bm{k}_1-\bm{k}_2|)}.
\label{equC1}
\end{equation}
To simplify the final expressions, we introduce in the integral the following notations:
\begin{equation}
  x=k_F^n+q,~~y=k_F^p-q,~~s=k_F^n-q,~~t=k_F^p+q.
\label{equC2}
\end{equation}
The final expression is then

\subsection{$k_F^n \geqslant 3k_F^p$ }

\begin{equation}
  P_{\mbox{av}}^2=
  \begin{cases}
     \frac{3}{5}(k_{F}^p)^2+q^2, & k_{F}^p \geqslant q \geqslant 0 \\
    \displaystyle \frac{\frac{8}{5}qs^5+yt^5+sxt^4-s^5x+\frac{2}{3}s^6-\frac{4}{3}t^6}
    {\frac{8}{3}qs^3+2yt^3+2sxt^2-2s^3x+s^4-2t^4}, & (k_{F}^n-k_{F}^p)/2 \geqslant q \geqslant k_{F}^p \\
    \displaystyle \frac{\frac{8}{5}q(s^5+y^5)+\frac{1}{12}(ty+sx)^3+\frac{2}{3}(s^6+y^6)
    -(ty^5+xs^5) }{\frac{8}{3}q(s^3+y^3)+\frac{1}{2}(ty+sx)^2+(s^4+y^4)-2(ty^3+xs^3)}, & (k_{F}^n+k_{F}^p)/2 \geqslant q > (k_{F}^n-k_{F}^p)/2\\
    0,& q > (k_{F}^n+k_{F}^p)/2.
  \end{cases}
\label{equC3}
\end{equation}

\subsection{$k_F^n \leqslant 3k_F^p$ }

\begin{equation}
  P_{\mbox{av}}^2=
  \begin{cases}
     \frac{3}{5}(k_{F}^p)^2+q^2, & (k_{F}^n-k_{F}^p)/2 \geqslant q \geqslant 0   \\
    \displaystyle \frac{\frac{8}{5}q(s^5+y^5)+\frac{1}{12}(ty+sx)^3+\frac{2}{3}(s^6+y^6)-(ty^5+xs^5) }{\frac{8}{3}q(s^3+y^3)+\frac{1}{2}(ty+sx)^2+(s^4+y^4)-2(ty^3+xs^3)}, & (k_{F}^n+k_{F}^p)/2 \geqslant q > (k_{F}^n-k_{F}^p)/2\\
    0,& q > (k_{F}^n+k_{F}^p)/2.
  \end{cases}
  \label{equC4}
\end{equation}

\end{appendix}


\end{document}